\documentclass{ws-procs975x65}

\begin{document}

\title{MULTIHORIZON REGULAR BLACK HOLES}

\author{S. ANSOLDI}

\address{I.C.R.A, I.N.F.N, and University of Udine,\\
via delle Scienze 206, Udine, I-33100, Italy\\
E-mail: ansoldi@fulbrightmail.org}

\author{L. SINDONI}

\address{Max Planck Institute for Gravitational Physics (Albert Einstein Institute),\\
Am M\"{u}hlenberg 1, Golm, D-14476, Germany\\
E-mail: Lorenzo.Sindoni@aei.mpg.de}

\begin{abstract}
We discuss a general procedure to generate a class of (everywhere regular) solutions of Einstein equations that can have an (a-priori fixed) arbitrary number of horizons. We then report on work currently in progress i) to find a suitable classification scheme for the maximal extension of these solutions and ii) to interpret the source term in Einstein equations as an effective contribution arising from higher dimensional and/or modified gravity.
\end{abstract}

\vskip 5 mm
\bodymatter
\noindent{}The study of the properties and structure of solutions to Einstein equations is a fascinating field: this is witnessed, for instance, by the history of the understanding of the first solution to these equations, the Schwarzschild solution \cite{bib:schsol}, whose analytical extension required more than half a century to be finally set on a rock solid ground.

In this contribution we show how it is possible to obtain \emph{non-vacuum}, spherically symmetric solutions that posses more than two horizons. Their general structure can be interpreted as a nested sequence of anisotropic radial vacua, with necessarily decreasing energy density. Although these solutions can be understood as \emph{black holes with regular center} \cite{bib:ansrev}, we will take a different point of view, and start from an apparently distinct idea, \emph{gravastars} \cite{bib:emimot}: common to both is the absence of singularities in the spacetime structure (achieved with a ``mild'' violation of energy conditions), but clearly a different attitude is taken with respect to the presence of horizons.

Let us then start from the neat analysis of Ref.~\citen{bib:matvis}, and consider the most generic line element for a static, spherically symmetric spacetime. Let us call $(t,r,\vartheta,\varphi)$ coordinates in which the metric components can be taken as
$
    g _{a b} = \mathrm{diag} ( \psi (r) , f ^{-1} (r) , r ^{2} , r ^{2} \sin ^{2} \vartheta )
$.
Let us moreover consider an energy-matter content that, in these very same coordinates, can be described by the stress-energy tensor
$
    T ^{a} {}_{b} = \mathrm{diag} ( - \rho (r) , p (r) , p _{\mathrm{t}} (r) , p _{\mathrm{t}} (r) ) ,
$
where $T ^{\vartheta} {}_{\vartheta} = T ^{\phi} {}_{\phi}$ is a consequence of spherical symmetry, but we allow for anisotropic effects to be present, $T ^{r} {}_{r} \neq T ^{\vartheta} {}_{\vartheta}$, and, for the moment, we also keep $T ^{t} {}_{t} \neq T ^{r} {}_{r}$.
If with foresight we anticipate the standard definition $m (r) = 4 \pi \int _{0} ^{r} \rho ( x ) x ^{2} dx$, then we fix boundary conditions so that $f (r) = 1 - 2 m (r) / r$; moreover, again following Ref.~\citen{bib:matvis}, we can express $\psi$ as\footnote{The functional dependence form the radial coordinate will be conveniently dropped after the following definition. Moreover, a prime will denote the first derivative with respect to $r$.}
\[
    \psi (r) = - \exp \left( - 2 \int _{r} ^{+ \infty} g (x) d x \right)
    , \; \mbox{so that we have} \quad
    g = \frac{m + 4 \pi p _{r} r ^{3}}{r ^{2} - 2 m r}
    ,
\]
as a consequence of one of the remaining, non-trivially satisfied, Einstein equations. The last of them, can be replaced with one of the energy conservation equations, $p ' _{r} = - ( \rho + p _{r}) g + 2 r ^{-1} ( p _{\mathrm{t}} - p _{r} )$. Let us now assume that the weak energy condition is satisfied everywhere ($\rho \geq 0$, $\rho + p _{r} \geq 0$ and $\rho + p _{\mathrm{t}} \geq 0$), and that the energy-matter content realizes what is called a radial vacuum \cite{bib:gli}, i.e. $p _{r} = - \rho$ (this identically satisfies the second of the weak-energy condition inequalities and implies $\psi f = - 1$). Then $- \rho ' = p ' _{r} = 2 r ^{-1} ( p _{\mathrm{t}} - p _{r} ) = 2 r ^{-1} ( p _{\mathrm{t}} + \rho ) \geq 0$ and we see that the energy density must be a \emph{non-increasing} function of $r$. Assuming a finite total mass energy $M = \lim _{x \to + \infty} m ( x )$ implies $\lim _{x \to + \infty} f ( x ) = 1$. Regularity of $\rho$ at the origin gives $\lim _{x \to 0 ^{+}} m ( x ) / x ^{3} = C \geq 0$; a solution which is regular there and globally non-trivial (not Minkowski) has $C > 0$, i.e. it is locally de Sitter around the origin.

\begin{figure}
\begin{center}
%\fbox{\vbox{%
\includegraphics[width=12.7cm]{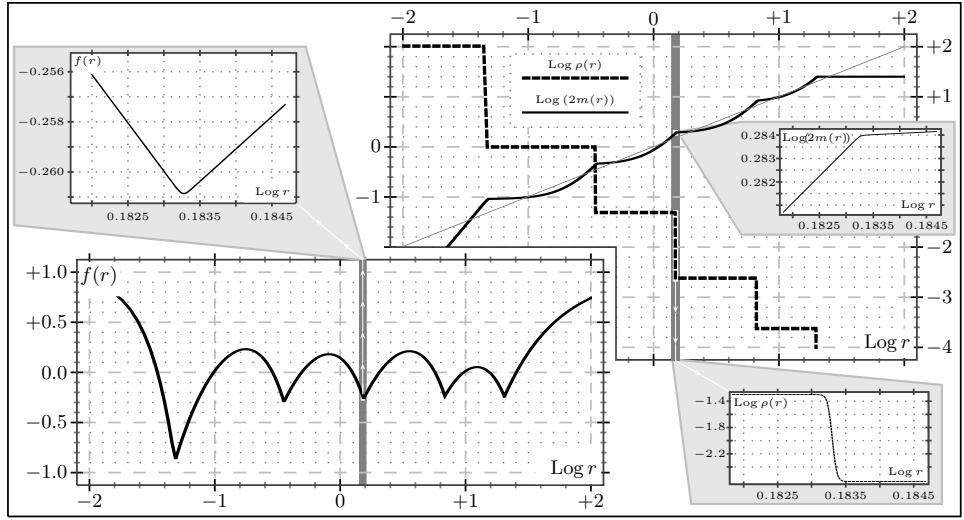}%
\caption[]{\label{fig:parsol}{\footnotesize{}%
Example of a solution with ten horizons of the kind described in the main text; the big top-right plot shows (the logarithms of) $\rho (r)$ (dashed line) and $m (r)$ (continuous line) \emph{versus} the logarithm of the radius ($\mathrm{Log} \equiv \log _{10}$). The bottom-left big panel shows the resulting $f (r)$. In the big panels the plot are clearly continuous, and the small panels give a visual intuition that all the functions are also smooth. This can also be seen analytically, as the function $\rho (r)$ is defined as:\newline
\centerline{$\displaystyle%
    \rho (r)
    =
    \left\{
        \begin{array}{ll}
            c _{i}
            &
            \quad \mbox{if} \quad a _{i - 1} \leq r < b _{i}
            \\[1mm]
            {%
                \displaystyle%
                \frac{c _{i} - c _{i+1}}{2}
                \left(
                    \mathrm{arctanh} \left[
                        \frac{(a _{i} + b _{i})(r - (a _{i} + b _{i}) / 2)}
                             {(r - a _{i})(r - b _{i})}
                    \right]
                    + 1
                \right)
                +
                c _{i}
            }
            &
            \quad \mbox{if} \quad b _{i} \leq r < a _{i}
            \\[1mm]
            0
            &
            \quad \mbox{if} \quad r \geq a _{5}
        \end{array}
    \right .
$}
where $i = 1 , 2 , \dots , 5$, $a _{0} = 0$, $\vec{a} = (0.05 , 0.355 , 1.55, 6.8, 20.5)$, $\vec{b} = (0.045, 0.35, 1.5, 6.75, 20)$, $\vec{c} = (100, 1, 0.05, 0.0025, 0.00025)$, $c _{6} = 0$ and we are using the notation $\vec{v} = (v _{1} , v _{2} , \dots , v _{5})$.
}}%}}
\end{center}
\end{figure}

Under these assumptions we now consider how many horizons one of these solutions can have and what kind of causal structures are allowed\footnote{Stability might be a serious concern for these solutions \cite{bib:masinf}, but in this contribution we are here merely concerned with their ``existence'' and we set this point aside.}. In the literature, all the examples of asymptotically flat black holes with a regular center known to the authors \cite{bib:ansrev,bib:brodym}, have a maximal extension with the causal structure of a Rei\ss{}ner-Nordstr\o{}m spacetime with a regular origin. However, we can immediately see that more general structures are also possible, as the equation $m = r$, or, more explicitly, $4 \pi \int _{0} ^{r} \rho ( x ) x ^{2} dx = r$, can have an arbitrary, even\footnote{When counted with the correct multiplicity.} number of solutions. Let us consider for $\rho$ a decreasing staircase function. Constancy of $\rho$ in a right neighborhood of the origin, implies that $m \sim r ^{3}$ will eventually equal $r$ and grow bigger. If now $\rho$ suddenly drops to a much lower value and then remains constant, we again have $m \sim r ^{3}$ but with a much lower cubic term coefficient. This can make $m$ grow slowly enough, until $r$ will become bigger than $m$ and we will have another zero of $f$. By tuning the length of the intervals in which $\rho$ remains constant and the corresponding values of $\rho$, we can generate an arbitrary number of zeroes of $f$. Of course, a discontinuous $\rho$ would result in $f$ being singular at the jumps. It is nevertheless possible to \emph{regularize} the jumps and obtain smooth $\rho$ and $f$ (see, e.g., Fig.~\ref{fig:parsol}). Following Ref.~\citen{bib:maxext}, it can be proved that in this case the resulting metric can be analytically continued across the zeros of $f$, and is thus regular there; possible peaks in $p _{\mathrm{t}}$ can be increasingly high but remain finite (this can be seen using $- \rho ' = 2 r ^{-1} ( p _{\mathrm{t}} + \rho )$). Already with four horizons, the Penrose diagram describing the maximal extensions is not planar, and several parts of the diagram may have multiple overlapping with other ones. This feature becomes more and more marked as the number of horizons grows. A classification algorithm of the resulting maximal extensions based on topological properties is under study and will be reported elsewhere \cite{bib:newpap}. Already with the example of Fig.~\ref{fig:parsol} it is nevertheless possible to have intuition and understanding about the existence of causal structures that have not yet been considered. As we will discuss elsewhere \cite{bib:newpap}, it is also suggestive to interpret these solutions as coming from higher dimensional and higher order pure gravity models.
\vspace*{-5mm}

\end{document}